\documentclass[a4paper]{jpconf}
\usepackage{graphicx}
\usepackage{hyperref}
\newcommand{\be}{\begin{equation}}
\newcommand{\ee}{\end{equation}}
\newcommand{\bea}{\begin{eqnarray}}
\newcommand{\eea}{\end{eqnarray}}
\def\eqa{\!\!&=&\!\!}
\def\ccr{\nonumber\\}

\def\bchi{\bar{\chi}}
\def\bpsi{\bar{\psi}}

\begin{document}
\title{Spinning particles and higher spin field equations}

\author{Fiorenzo Bastianelli$^1$, Roberto Bonezzi$^1$, Olindo Corradini$^{2,3}$ and  Emanuele Latini$^4$}

\address{$^1$ Dipartimento di Fisica ed Astronomia, Universit{\`a} di Bologna and\\
INFN, Sezione di Bologna, via Irnerio 46, I-40126 Bologna, Italy\\
$^2$ Facultad de Ciencias en F\'isica y Matem\'aticas,
Universidad Aut\'onoma de Chiapas, Ciudad Universitaria, Tuxtla Guti\'errez 29050, M\'exico\\
$^3$ Dipartimento di Scienze Fisiche, Informatiche e Matematiche\\ Universit\`a di Modena e Reggio Emilia, Via Campi 213/A, I-41125 Modena, Italy\\
$^4$ Institut f{\"u}r Mathematik, Universit{\"a}t Z{\"u}rich-Irchel, Winterthurerstrasse 190, CH-8057 Z{\"u}rich, Switzerland}

\begin{abstract}
Relativistic particles with higher spin can be described in first quantization using actions with local supersymmetry on the worldline.
First, we present a brief review of these actions and their use in first quantization.
In a Dirac quantization scheme the field equations emerge as Dirac constraints on the Hilbert space,
and we outline how they lead to the description of higher spin fields in terms of the more standard Fronsdal-Labastida equations.
Then,  we describe how these actions can be extended so that the propagating particle is allowed to take different values of the spin,
i.e. carry a reducible representation of the Poincar\'{e} group. This way one may
identify a four dimensional model that carries the same degrees of freedom of the minimal
Vasiliev's interacting higher spin field theory.
Extensions to massive particles and to propagation on (A)dS spaces are also briefly commented upon.
 \end{abstract}

\section{Introduction}
Spinning particle actions based on local supersymmetry on the worldline \cite{Gershun:1979fb,Howe:1988ft}
identify amusing systems that exemplify several theoretical constructions and form an arena where to test various methods and ideas.
From this perspective they attracted the interest of Victor, who wanted to use them to test quantization methods whose development
he had contributed to \cite{Villanueva:1999st}. These models can be employed to study higher spin field theories in first quantization.

Higher spin theories have recently attracted much interest in a desire to understand better
Vasiliev's constructions of interacting theories of higher spin fields \cite{Vasiliev:1990en, Vasiliev:2003ev}
and their use in AdS/CFT dualites \cite{Sezgin:2002rt, Klebanov:2002ja}.
Vasiliev's theories are non-lagrangian, and it is at present  unclear how to perform their quantization.
A first quantized approach, similar to the one used in string theory, might be welcome to address the problem.
With this perspective in mind, we have analyzed first quantization of the (higher) spinning particles
in a series of paper  \cite{Bastianelli:2007pv, Bastianelli:2008nm, Bastianelli:2012bn, Bastianelli:2014lia},
whose content will be summarized in the following. After that we discuss how to modify the action to allow the
particle to take different values of the spin, i.e. carry a reducible representation of the Poincar\`e group.
Ideally, one would like to construct a system that carries the same
degrees of freedoms of the  Vasiliev's  theories, be it a particle, string or more general mechanical system, as a first step
towards a first quantized realization of interacting higher spin theories. Here we will find a particle model that carries the same degrees of freedom
of a four dimensional Vasiliev's  theory, though the model is constructed in flat space and extension to AdS remains unclear.

\section{Actions}
The class of higher spin particles that we are going to discuss are singled out
by actions with an $O(N)$-extended supersymmetry on the worldline.
They are constructed as supersymmetric extensions of the usual relativistic scalar particle action.
The latter has  a geometrical interpretation: it is proportional to the length of the worldline.
In natural units it is given by
\be
S=-m \int ds
\label{bos-act}
\ee
where $ds=\sqrt{-\dot x^\mu \dot x_\mu}$, with $x^\mu(\tau)$ the functions that describe the worldlines
in a flat spacetime with cartesian coordinates $x^\mu$. The parameter $\tau$ is largely arbitrary, as one can perform
reparametrizations of the form
\be
\tau \to \tau'=\tau'(\tau) \;, \qquad  x^\mu(\tau) \to {x'}^\mu(\tau') = x^\mu(\tau)
\label{2}
\ee
as long as these transformations are invertible. This is a local symmetry that is crucial to
keep Lorentz invariance manifest and recover unitarity of the quantum theory.
Indeed one may use the freedom of selecting which parameter to use,
and choose $\tau=x^0$ so that $x^0$ stops being a dynamical variable. Then, the action takes the following
standard form without gauge symmetries
\be
S[{\bf x}(t)] = -m \int dt\ \sqrt{1-\dot {\bf x}(t) \cdot\dot {\bf x}(t)}
\ee
where $t\equiv x^0$  is the time and $\bf x$ the space  coordinates. However, it is often preferable
to keep a manifest Lorentz invariance, and accept a redundancy in the description of the system, by using all of the $x^\mu(\tau)$
as dynamical variables. The local symmetry in (\ref{2}) manifests itself in the form of a first class constraint when considering
 the hamiltonian formulation, a preliminary step for canonical quantization.
Computing the conjugate momenta  $p_\mu$ from (\ref{bos-act}), one finds a constraint, the  mass-shell constraint, and a vanishing canonical hamiltonian $H_c$
\be
p_\mu =\frac{ \partial L}{\partial \dot x^\mu} =\frac{m \dot x_\mu}{\sqrt{-\dot x^2}} \quad \to \quad p_\mu p^\mu + m^2 =0 \;,
\qquad \ \ H_{c}= p_\mu \dot x^\mu - L = 0 \;.
\ee
The nontrivial dynamics is fully contained in the constraint, traditionally called $H$ and  normalized as
\be
H\equiv \frac12 (p_\mu p^\mu + m^2) =0 \;.
\ee
 It generates the gauge transformations (the reparametrizations)
in phase space. The phase space action takes the form
\be
S[x^\mu, p_\mu, e] =\int d\tau\, ( p_\mu \dot x^\mu - e H)
\ee
where $e$ is the Lagrange multiplier (the einbein) that implements the constraint $H=0$. It is a gauge field, since under reparametrizations it transforms as the derivative of the infinitesimal gauge parameter $\zeta$
\be
\delta x^\mu = \zeta p^\mu \;, \qquad
\delta p_\mu = 0
\;, \qquad
\delta e = \dot \zeta  \;.
\ee
Eliminating the momenta  by their algebraic equations of motion $p^\mu = e^{-1}\dot x^\mu$,
one finds the configuration space action
\be
S[x^\mu, e] =  \int d\tau\ \frac{1}{2}
( e^{-1} \dot x^\mu \dot x_\mu - e m^2)
\ee
which has the advantage over (\ref{bos-act}) of having a smooth massless limit, just as the phase space action.
It can be put in an arbitrarily curved space, and in that form was used in \cite{Bastianelli:2002fv}
to develop a worldline approach to scalar fields coupled to background gravity.

The $N=1$ supersymmetric extension of the scalar model produces an action for a spin 1/2 particle.
One introduces fermionic partners $\psi^\mu$ to the bosonic coordinates $x^\mu$, and gauges the supersymmetry
that relates them. The action in phase space takes the form
 \bea
S \eqa \int d\tau
\left (
p_\mu \dot x^\mu + \frac{i}{2}  \psi_\mu \dot \psi^\mu  - e H - i \chi Q
\right )
\eea
where the first class constraints given by
\be
H= \frac12 p^2    \;, \qquad
Q= p_\mu \psi^\mu
\ee
are gauged by the einbein $e$ and gravitino $\chi$. Both the $\psi^\mu$'s and $\chi$ are real Grassmann valued variables, and $Q$ is called the susy charge
as it generates supersymmetry transformations on the worldline.
The constraints realize the $N=1$ susy algebra in one dimension  through the Poisson brackets
\be
\{Q,Q\}= -2i H \;.
\ee
Upon quantization the $\psi^\mu$ play the role of the gamma matrices, and the constraint
$Q=0$ becomes the massless Dirac equation in the Dirac quantization scheme.
A mass term can be introduced by dimensional reduction.
This action was formulated in \cite{Brink:1976sz}, and its quantum mechanics analysed in \cite{Henneaux:1982ma}.
It is used in \cite{Bastianelli:2002qw, Bastianelli:2003bg}
for developing a worldline description of quantum Dirac fields coupled to background gravity.

The extension to $N=2$ supersymmetries is quite instructive, as it contains additional elements useful for understanding the general case.
One introduces complex fermionic parterns $\psi^\mu$ and $\bar\psi^\mu$
to the coordinates $x^\mu$, and gauges the full $N=2$ extended supersymmetry
that relates them. The action in phase space takes the form
\be
S  = \int d\tau
\left (  p_\mu \dot x^\mu + i \bpsi_\mu \dot \psi^\mu -e H - i \bchi Q - i \chi \bar Q - a (J-c) \right )
\label{n=2}
\ee
where the first class constraints
\bea
H = {1\over 2} p_\mu p^\mu \ , \quad
Q = p_\mu \psi^\mu \ , \quad
\bar Q = p_\mu \bar \psi^\mu \ , \quad
J =  \bar \psi^\mu \psi_\mu  \ .
\eea
are gauged by the variables $e, \bar \chi, \chi, a$.
The constraints realize the $N=2$ extended susy algebra
\bea
\{Q,\bar Q\} = -2i H \ , \quad
\{J,Q\} = i Q  \ , \quad
\{J,\bar Q\} = - i \bar Q
\label{6.algebra}
\eea
(other Poisson brackets vanish).  Note that there is a $U(1) \sim SO(2)$ group,
the so-called $R$-symmetry group, generated by the charge $J$
that is gauged by the $U(1)$ gauge field $a$. The latter is allowed to have an additional Chern-Simons coupling constant $c$, whose
net effect is to modify the constraint implemented by $a$  from $J =0$ to $J-c =0$.
The action is manifestly Poincar\'e invariant in target space and thus identifies a relativistic model.
The $H, Q,\bar Q$ constraints guarantee unitarity, as they can be used to eliminate the negative norm states generated by the variables
$x^0, \psi^0, \bpsi^0$,  while the $J$ constraint guarantees
irreducibility of the model, i.e. it describes a particle that carries an irreducible representation of the Poincar\'e group of target space.
It is seen that this model describes a spin 1 massless particle through the free Maxwell equations.
Let us explain how this arises in some detail, considering a spacetime of dimension $D=4$ for simplicity.
 Wave functions can be seen as depending on the generalized coordinates $x^\mu$ and $\psi^\mu$
\be
\phi(x,\psi) = F(x) + F_\mu(x) \psi^\mu +
{1\over 2} F_{\mu\nu}(x) \psi^{\mu}\psi^{\nu} +
{1\over 3!} F_{\mu\nu\rho}(x) \psi^{\mu}\psi^{\nu}\psi^{\rho} +
{1\over 4!} F_{\mu\nu\rho\sigma}(x) \psi^{\mu}\psi^{\nu}\psi^{\rho} \psi^{\sigma}
\label{sviluppo}
\ee
whose Taylor expansion in the $\psi$'s stops, as the latter are Grassmann variables.
The momenta  $p_\mu=-i \frac{\partial\ }{\partial x^\mu}$ and $ \bpsi_\mu=\frac{\partial_L\ }{\partial \psi^\mu}$ act as derivatives
(we use left derivatives for the Grassmann variables, meaning that we remove the increment form the left).
Now, the classical constraints $C$ become differential operators $\hat C$
that select physical wave functions by requiring  $\hat C\phi_{phys}(x,\psi)=0$. The $J$ constraint suffers from quantum ordering ambiguities,
so that using the antisymmetric ordering $J= \frac12(\bar \psi^\mu \psi_\mu  -\psi_\mu  \bar \psi^\mu)$
one finds the differential operator $\hat J =2- \psi^\mu  \frac{\partial_L\ }{\partial \psi^\mu}$.
Choosing a vanishing Chern-Simons coupling, one finds
the constraint $\hat J\phi_{phys}(x,\psi)=0$ which is solved by
\be
\phi_{phys} (x,\psi) =
{1\over 2} F_{\mu\nu}(x) \psi^{\mu}\psi^{\nu} \;.
\ee
Then, the constraints $\hat Q \phi_{phys} = 0$ gives integrability conditions on the surviving tensor
$F_{\mu\nu}(x)$  (Bianchi identities upon the introduction of a gauge potential)
\be
\partial_{\rho} F_{\mu\nu}+ \partial_{\mu} F_{\nu\rho}+ \partial_{\nu} F_{\rho\mu}
 =0
\ee
and the constraint $\hat Q^\dagger \phi_{phys} =0 $,  arising form quantizing $\bar Q$, produces the remaining free Maxwell equations
\be
\partial^{\mu} F_{\mu\nu} =0 \ .
\ee
Thus we see how the standard description of a free spin 1 massless particle emerges in first quantization.
More generally, one can use different values of the quantized Chern-Simons coupling to describe differential $p$-forms satisfying generalized Maxwell equations
in arbitrary dimensions \cite{Howe:1989vn}.
This description was used in \cite{Bastianelli:2005vk, Bastianelli:2005uy} to treat spin 1 and antisymmetric tensor fields coupled to gravity in first quantization.

The general case, where one introduces $N$ real fermionic partners $\psi_i^\mu$ ($i=1,..., N$) associated to the bosonic
coordinates $x^\mu$, and gauges the resulting $O(N)$-extended supersymmetry present on the worldline,
was discussed in \cite{Gershun:1979fb, Howe:1988ft}, and describes a particle of spin $s=\frac{N}{2}$. The action in phase space takes the form
\bea
S \eqa \int d\tau
\left ( p_\mu \dot x^\mu + {i \over 2} \psi_{i\mu} \dot \psi_i^\mu
-e H
- i \chi_i Q_i
- {1 \over 2} a_{ij} J_{ij}
\right )
\eea
where the first class constraints
\be
H = \frac12 p_\mu p^\mu \;, \quad
Q_i =  p_\mu \psi^\mu_i \;, \quad
J_{ij} = i \psi_i^\mu \psi_{j\mu}
\ee
are gauged by the fields $e,\chi_i,a_{ij}$.
The Poisson bracket algebra of the constraints is indeed that of the $O(N)$-extended supersymmetry
\bea
\{Q_i,Q_j\} \eqa -2i \delta_{ij} H \;, \qquad
\{J_{ij},Q_k\} = \delta_{jk} Q_i -\delta_{ik} Q_j  \ccr[1mm]
 \{J_{ij},J_{kl}\}  \eqa \delta_{jk} J_{il} - \delta_{ik} J_{jl}
- \delta_{jl} J_{ik} + \delta_{il} J_{jk}  \;.
\label{ca}
\eea
Quantization \`a la Dirac shows that the model describes a massless particle with spin $s=\frac{N}{2} $ in terms of the Bargmann-Wigner equations
\cite{Bargmann:1948ck}. Let us review briefly the analysis for integer spin $s$. In this case one can form complex combinations
of the fermionic variables
\be
\psi_I^\mu = \frac{1}{\sqrt{2}}(\psi_i^\mu + i \psi_{i+s}^\mu) \;, \qquad
\bar \psi^{\mu I} =\frac{1}{\sqrt{2}}(\psi_i^\mu - i \psi_{i+s}^\mu) \;, \qquad \ \ I=i=1,..,s
\ee
and have $s$ pairs (indexed by $I$) of fermionic creation/annihilation operators
(of course, each pair has an additional spacetime index acting as a spectator).
In this basis only the subgroup $U(s) \subset SO(2s)$ is manifest.
As in the $N=2$ case, one can take the $\psi_I^\mu$ as fermionic coordinates, on which a generic wave function may depend,
and  $\bar \psi^I_\mu = \frac{\partial\ }{\partial \psi_I^\mu}$
 as corresponding momenta, realized as left derivatives with respect to the coordinates.
Then it follow that the generic wave function $R(x,\psi)$ contains all possible tensors having $s$ blocks of antisymmetric indices,
as the indices of each block arise from the Taylor expansion of the same type of fermion, (i.e. a fermion with the same internal index $I$),
in a way similar to  what seen in eq. (\ref{sviluppo}). Then one must impose the quantum constraints on the wave function.
The constraints due to the $SO(N)$ charges $J_{ij}$ select the tensor
with $s$ blocks of $d=\frac{D}{2} $ indices, implying that a nontrivial solution is present for even spacetime dimensions $D$ only
\be
R_{\mu^1_1..\mu^1_d, ..., \mu^s_1..\mu^s_d} \;.
\label{23}
\ee
In addition, the $J_{ij}$ constraints require this tensor to
be totally traceless and with the symmetries of a Young tableau with $d$ rows and $s$ columns.
The susy charges can also be split in pairs of complex conjugates charges, just like the fermions.
The contraints from the susy charges $Q_I =  p_\mu \psi^\mu_I $
imply integrability conditions of the form
\be\label{closed}
\partial_{[\mu} R_{\mu^1_1..\mu^1_d], ..., \mu^s_1..\mu^s_d} =0
\ee
for each block (interpreted as Bianchi identities once solved).
The other half of susy charges  $\bar Q_I =  p_\mu \bar \psi^\mu_I $ produce
``Maxwell equations'' of the form
\be
\partial^{\mu} R_{\mu \mu^1_2..\mu^1_d, ..., \mu^s_1..\mu^s_d} =0  \ .
\ee
The $ H$ constraint is satisfied identically as consequence of the algebra.
These are the geometric equations that describe a free field of spin $s$,
equivalent to the massless Bargmann-Wigner equations,
usually given with the wave function in  a multispinor basis  \cite{Bargmann:1948ck}.
They are called geometric as the tensors $R$ can be interpreted as linearized curvatures.
A proper analysis of these equations, showing in particular that they are equivalent to Fronsdal-Labastida ones and that they propagate the correct degrees of freedom,
has been carried out in \cite{Bekaert:2002dt, Bekaert:2003az, Bekaert:2003zq}, also reviewed in \cite{Bekaert:2006ix}.
The above equations have  also the property of being conformal invariant \cite{Siegel:1988ru,Siegel:1988gd}.
They can be related to the more standard description in terms of the Fronsdal-Labastida equations
\cite{Fronsdal:1978rb, Labastida:1987kw}, as we are going to describe next.

\section{Fronsdal-Labastida equations}
In this section we discuss how the Fronsdal-Labastida equations, describing the propagation of a massless
particle of integer spin $s=\frac N 2$, emerge
from solving some of the constraints of the canonical analysis of the previous section.
In analogy with the case of the electromagnetism, we introduce the gauge potential
\be\label{varphi}
\varphi(x,\psi):= \varphi(x)_{\mu^1_1..\mu^1_{d-1}, ..., \mu^s_1..\mu^s_{d-1}}\psi^{\mu^1_1}_1..\psi^{\mu^1_{d-1}}_1\cdots \psi^{\mu^s_1}_s..\psi^{\mu^s_{d-1}}_s
\ee
with the symmetries of a rectangular Young tableau with $d-1$ rows and $s$ columns,
and we solve the condition (\ref{closed}) by setting
\be
R(x,\psi)=Q_1\cdots Q_s \varphi(x,\psi)
\ee
where we have used the complex charges $Q_I$ with $I=1,...s$.
A generalisation of the Poincar{\'e} lemma assures that this solution is unique, modulo the equivalence relation, or gauge symmetry,  of the form
$\varphi(x,\psi)\sim \varphi(x,\psi)+Q_I \xi^I(x,\psi)$. Observe now that the curvature must vanish when we trace over the sectors $I$ and $J$,
making the tensor in (\ref{23}) totally traceless:
 this condition is encoded into the $SO(N)$ constraint $J^{IJ}$ and produces the following equation
\be\label{qsequation}
\frac{\partial}{\partial \psi_I^\rho}\frac{\partial}{\partial \psi_{J \rho} } R(x,\psi)=0 \quad \,\,\,\rightarrow\quad \,\,\,Q_1..Q_{I-1}Q_{I+1}..Q_{J-1}Q_{J+1}..Q_s({\bf G} \varphi(x,\psi) )=0
\ee
where we have introduced the Fronsdal-Labastida operator defined by
\be
{\bf G} := -2H+Q_I\bar{Q}^I
+\frac 1 2 Q_IQ_J J^{IJ} \;.
\ee
The next task is to get rid of the operator on the left of the Fronsdal-Labastida operator in (\ref{qsequation}). This can be achieved by introducing a new independent field, called the compensator
\be
\rho^{IJK}(x,\psi):=w^{\mu\nu\delta}(x)\frac{\partial}{\partial \psi_I^\mu}\frac{\partial}{\partial \psi_{J }^{\nu} }\frac{\partial}{\partial \psi_K^\delta}\rho(x,\psi)
\ee
where $\rho(x,\psi)$ has the symmetries and expansion of the gauge potential given in (\ref{varphi}).
The compensator parametrizes elements in the kernel of the operator in (\ref{qsequation}) acting on $ {\bf G} \varphi(x,\psi) $.
Thus, one may rewrite eq. (\ref{qsequation}) as
\be
{\bf G} \varphi(x,\psi)=Q_IQ_JQ_K \rho^{IJK}(x,\psi) \;.
\ee
This is the Fronsdal-Labastida equation with compensator. The latter was
introduced in this context in \cite{Francia:2002pt}.
One can also use part of the gauge symmetry to set the compensator to zero, and recover
the original Fronsdal-Labastida system
\be
{\bf G}\, \varphi(x,\psi)=0
\label{FLeq}
\ee
that enjoys the residual gauge invariance with a traceless gauge parameter
\be
\delta \varphi = Q_{I}\xi^{I}(x,\psi)\,,\,\,\,\,\textrm{with}\,\,\,\, J^{[IJ}\xi^{K]}(x,\psi)=0\;.
\ee
Note that, for consistency, the gauge field has to be doubly traceless $J^{IJ}J^{KL}\varphi(x,\psi)=0$.
This condition is a consequences of (\ref{FLeq}).
In four dimensions the gauge field is a completely symmetric tensor $\varphi_{\mu_1\cdots \mu_s}$, and eq. (\ref{FLeq})
translates into the Fronsdal equation
\be
\partial^\nu\partial_\nu\varphi_{\mu_1\cdots \mu_s} - (\partial_{\mu_1}\partial^{\nu}\varphi_{\nu\mu_2\cdots\mu_s}+\cdots)+(\partial_{\mu_1}\varphi^{\nu}_{\,\,\,\,\nu\mu_3\cdots_{\mu_s}}+\cdots)=0
\ee
where the brackets contain $s$ and $\frac1 2 s(s-1)$ terms, respectively, needed for symmetrizing the $\mu$ indices.
Of course, the gauge field must be doubly traceless, i.e.  $  \varphi^{\nu\delta}_{\,\,\,\,\,\,\nu\delta\mu_5\cdots \mu_s}=0$.

We have presented the analysis for integer spins only, but a similar program can be carried out  for
the case of half-integer spins as well \cite{Corradini:2010ia}.
The equivalent BRST quantization for this model is described in  \cite{Marnelius:1988ab, Siegel:1999ew}.
In particular, in \cite{Siegel:1999ew} one finds its use to construct second quantized actions for any spin in flat spaces of arbitrary dimensions.
Quantized point particles of any spin have been treated also in \cite{Henneaux:1987cp}.

\section{Other gaugings and the case of $U(s)$}

We have seen that the constraints of the $O(N)$ spinning particle allow to recover a unitary irrep of the Poincar\'e group,
corresponding to a massless particle with fixed spin (helicity). Unitarity is guaranteed by the hamiltonian constraint $H$
and the susy constraints $Q_i$, as they can be used to remove the dangerous polarizations in the wave function
generated by the $x^0$ and $\psi^0_i$ quantum variables. The additional constraints due to the $O(N)$ charges $J_{ij}$
can be relaxed without destroying unitarity of the quantum theory, as they serve the only purpose of selecting an
irreducible representation of the Poincar\'e group. In fact, one may actually prefer to describe the propagation
of a multiplet of states, especially if one imagines that they can be made interacting  somehow.
One example was treated in \cite{Pashnev:1990cf} for the $O(4)$ particle in $D=4$. As $O(4)\sim SU(2)\otimes SU(2)$,
Pashnev and Sorokin gauged only a $SU(2)$ factor, finding that the emerging model propagates a graviton
and three scalars. In general,  one may investigate the consequences of gauging  different subgroups of the $O(N)$ symmetry group.
For example,
one may not gauge anything at all, and thus find the propagations of a maximum number of states with different spins.
A more refined option is to gauge the $U(1)^s$ subgroup of $O(N)$ (we consider even $N=2s$, restricting  ourselves to bosonic particles).
Each factor $U(1)$  may have an additional independent  Chern-Simons coupling, useful to project to the subsector of the wave functions
 containing a fixed number of antisymmetric indices of each favour (different $I$'s indicate different flavours).
  This projection was briefly discussed earlier for the $N=2$ case, and constitutes a
 useful trick to project to a desired subsector of the Hilbert space. It has been used often in worldline applications, as in
  \cite{Bastianelli:2013pta} and \cite{Bastianelli:2013tsa}.

 In this section we are going to analyze the gauging of a $U(s)$ subgroup of the full $R$-symmetry group $O(2s)$. In the previous sections the $O(2s)$ generators were split in $U(s)$ covariant form as $J_{ij}=(J_{IJ}, J^{IJ}, J^I_J)$, where $J_{IJ}$ and $J^{IJ}$ insert the metric tensor and compute a trace in the $IJ$ family of indices, respectively, while $J^I_J$ is the $U(s)$ generator that performs anti-symmetrization of indices between the aforementioned families (we recall that an upper index $J$ is equivalent to a lower index $\bar J$).
  By gauging only the $J^I_J$ generators one gains the freedom of adding a Chern-Simons coupling $c$ as in the $N=2$ model, see eq. (\ref{n=2}). The phase space action will thus read
\be
S  = \int d\tau
\left (  p_\mu \dot x^\mu + i \bpsi^I_\mu \dot \psi_I^\mu -e H - i \bchi^I Q_I - i \chi_I \bar Q^I - a^I_J (J^J_I-c\,\delta_I^J) \right )\;.
\label{Us}
\ee
The presence of the Chern-Simons coupling allows to set the eigenvalue of the number operators
$N_I:=J^I_I$ with fixed $I$ (i.e. no summation)
to any desired value. In particular, by setting $c=p+1-D/2$ one obtains curvature tensors described by rectangular Young tableaux with $p+1$ rows and $s$ columns in any spacetime dimension $D$. Indeed, the covariant Dirac quantization of the model gives as physical field a curvature $R$, which forms an irreducible $GL(D)$ tensor with the symmetries of the rectangular Young tableau described above, obeying the Maxwell-like equations
\begin{equation}\label{Maxwell}
Q_I\,R=\bar Q^I\,R=0\;.
\end{equation}
These Maxwell-like fields propagate a multiplet of single-particle states, as the trace constraint is not imposed.
This is analogous to the case of the $U(N)$ spinning particles, introduced in \cite{Marcus:1994mm} and analyzed in
\cite{Bastianelli:2009vj}, modelling particles in a euclidean complex space, which do not have any natural trace operator
(see refs. \cite{Bastianelli:2011pe, Bastianelli:2012nh} for further analysis of these amusing systems).
As field theory models, these  Maxwell-like fields were introduced at the level of gauge potentials in \cite{Campoleoni:2012th} following ideas developed in \cite{Francia:2010qp}, while their curvature description was studied in \cite{Francia:2012rg, Bekaert:2015fwa}.

In the following we perform a light-cone analyses of the particle model in order to count the physical degrees of freedom.
We define light-cone coordinates in target space as $x^{\pm}:=\frac{1}{\sqrt2}(x^1\mp x^0)$, while $x^a$ denote transverse coordinates, such that $A\cdot B=A_+B_-+A_-B_++A_a B_a$. By using worldline reparametrizations one can gauge fix $x_+=\tau$ and solve the corresponding constraint $H=0$ by setting $p_-=-\frac{p_ap_a}{2p_+}$, since $p_+$ is assumed to be invertible in light-cone analysis. Now one can use the local supersymmetries to gauge fix $\psi_{+I}=\bar\psi_+^I=0$, and solve $Q_I=\bar Q^I=0$ with $\psi_{-I}=-\frac{p_a\psi_{aI}}{p_+}$ and $\bar\psi_{-}^I=-\frac{p_a\bar\psi_{a}^I}{p_+}$. The action thus reduces to
\be
S  = \int d\tau
\left (  p_+ \dot x^++p_a\dot x^a-\frac{p_ap_a}{2p_+} + i \bpsi^I_a \dot \psi_I^a - a^I_J (\hat J^J_I-c\,\delta_I^J) \right )\;,
\label{lightcone}
\ee
where we have defined the reduced $U(s)$ generator as $\hat J^I_J=\psi_J^a\,\bar\psi^I_a$. At the quantum level this generator suffers from
an ordering ambiguity, for which we choose again the anti-symmetric ordering, such that
$$
\hat J^I_J=\frac12[\psi_J^a,\bar\psi^I_a]=\psi_J^a\,\bar\psi^I_a-\delta^I_J\,\frac{D-2}{2}\;,
$$
and the quantum constraint $\hat J_J^I-c\,\delta^I_J$ acts as the differential operator $(\psi^a_I\frac{\partial}{\partial\psi^a_J}-p\,\delta_I^J)$, for $c=p+1-\frac{D}{2}$. One can see then that the states are tensors of $GL(D-2)$, and the constraint $\hat J^I_J-c\,\delta^I_J$ provides $GL(D-2)$ irreducibility by fixing the length of the columns to be $p$, and by gluing the columns together in a rectangular Young tableau with $p$ rows and $s$ columns.
The physical state consists thus of a massless irreducible tensor of $GL(D-2)$ with the symmetries spelled above. Its particle spectrum is  provided by the branching of the rectangular Young tableau of $GL(D-2)$ into traceless $SO(D-2)$ representations. In particular, for $p=1$ we have a symmetric $GL(D-2)$ tensor of rank $s$, and the spectrum is given by massless particles of spin $s,s-2,s-4,..$ down to spin zero or one,
according to if $s$ is even or odd, respectively.

 It is rather interesting that taking even spin $s$ and eventually sending $s\to\infty$ one finds the spectrum of the Vasiliev's minimal bosonic models, that contain even spins ranging from zero to infinity \cite{Vasiliev:2003ev}.


\section{Massive particles and  couplings to AdS}

A class of actions for massive particles of higher spins can be obtained by dimensionally reducing the massless model
through the  Scherk-Schwarz mechanism \cite{Scherk:1979zr}.
The massive particle lives in odd dimensions, as the original massless model with the fully gauged $O(N)$-extended
supersymmetry lives in  even dimensions.  Further, taking the massless limit produces a model propagating
several helicities. This procedure gives another method for generating actions for  the propagation of a multiplet
of particle states \cite{Bastianelli:2014lia}. In addition, one might add the option of gauging a smaller subgroup
 of the $R$-symmetry group, thus producing an extended range of possibilities.
 We are not going to analyze them further here.

A different option to construct worldline models for higher spinning particles is  that of gauging some part of the symmetry algebra
of the $Sp(2N)$ quantum mechanics described in \cite{Hallowell:2007zb}, where instead of anticommuting variables $\psi^\mu_i$
one uses complex commuting variables $z^\mu_i$ and  $\bar z^\mu_i$. This approach has been analyzed in \cite{Bastianelli:2009eh}.

 A key problem is to study allowed interactions of massless particles of higher spin. As already mentioned, a set of consistent interacting
 models has been produced by Vasiliev, that constructed suitable nonlinear field equations on (A)dS backgrounds.
 At the first quantized level, an initial step it to study how the $O(N)$ spinning particles could be coupled to a nontrivial background.
 For a while it was thought that no coupling could be allowed at all, but eventually a way to couple them to the (A)dS spaces \cite{Kuzenko:1995mg},
 and more generally to conformally flat spaces \cite{Bastianelli:2008nm}, was found.
 In \cite{Bastianelli:2008nm} the method employed was that of covariantizing the  constraint algebra in phase space, and check if
 the algebra could be still of first class. On (A)dS the algebra becomes quadratic, with  the only modification with respect to
 the flat space case (\ref{ca}) sitting in the Poisson brackets of the susy charges
 \bea
\{Q_i,Q_j\} \eqa -2i \delta_{ij} H +ib \Big (J_{ik}J_{jk}
-\frac{1}{2}\delta_{ij} J_{kl} J_{kl}\Big)
\eea
 where the constant $b$ is related to the (A)dS curvature scalar by
$b=\frac{R}{D(D-1)}$.  On conformally flat backgrounds, the algebra acquires more complicated structure functions.
The model can be quantized, and in \cite{Bastianelli:2012bn}
we have used it to study the one-loop effective action on (A)dS spaces, computing the first few Seeley-DeWitt coefficients, namely
those ones that are related to the diverging terms in four dimensions. A similar calculation presumably could be carried out for the massive models
introduced in \cite{Bastianelli:2014lia}, though the gauge fixing procedure may  be more difficult because of the more complicated
structure functions. We expect that restricting the calculation of the effective action to the (A)dS spaces from the start
(in \cite{Bastianelli:2012bn} we have kept the background arbitrary, and restricted to the (A)dS  case only at the end)
one might push the calculation to the next perturbative order (a three loop calculation on the worldline).
The conterterms to be used for such a calculation are already available \cite{Bastianelli:2011cc}.
On an arbitrary geometry a  three loop calculation is quite demanding \cite{Bastianelli:2000dw}, but  the
simplications due to restricting to (A)dS spaces make it much more manageable \cite{Bastianelli:2001tb}.
We plan to report on that task in the near future.

\section*{Acknowledgments}{The work of OC was partly supported by the UCMEXUS-CONACYT grant CN-12-564. EL acknowledges partial support from SNF Grant No. 200020-149150/1.}

\section*{References}

\begin{thebibliography}{99}

\bibitem{Gershun:1979fb}
  V.~D.~Gershun and V.~I.~Tkach,
  ``Classical and quantum dynamics of particles with arbitrary spin,''
  JETP Lett.\  {\bf 29} (1979) 288
   [Pisma Zh.\ Eksp.\ Teor.\ Fiz.\  {\bf 29} (1979) 320].

\bibitem{Howe:1988ft}
  P.~S.~Howe, S.~Penati, M.~Pernici and P.~K.~Townsend,
  ``Wave equations for arbitrary spin from quantization of the extended supersymmetric spinning particle,''
  Phys.\ Lett.\ B {\bf 215} (1988) 555.

\bibitem{Villanueva:1999st}
  V.~M.~Villanueva, J.~Govaerts and J.~L.~Lucio-Martinez,
 ``Quantization without gauge fixing: avoiding Gribov ambiguities through the physical projector,''
  J.\ Phys.\ A {\bf 33} (2000) 4183
  [hep-th/9909033].

\bibitem{Vasiliev:1990en}
  M.~A.~Vasiliev,
  ``Consistent equation for interacting gauge fields of all spins in (3+1)-dimensions,''
  Phys.\ Lett.\ B {\bf 243} (1990) 378.

\bibitem{Vasiliev:2003ev}
  M.~A.~Vasiliev,
  ``Nonlinear equations for symmetric massless higher spin fields in (A)dS(d),''
  Phys.\ Lett.\ B {\bf 567} (2003) 139
  [hep-th/0304049].

\bibitem{Sezgin:2002rt}
  E.~Sezgin and P.~Sundell,
  ``Massless higher spins and holography,''
  Nucl.\ Phys.\ B {\bf 644} (2002) 303
   [Erratum-ibid.\ B {\bf 660} (2003) 403]
  [hep-th/0205131].

\bibitem{Klebanov:2002ja}
  I.~R.~Klebanov and A.~M.~Polyakov,
  ``AdS dual of the critical O(N) vector model,''
  Phys.\ Lett.\ B {\bf 550} (2002) 213
  [hep-th/0210114].

\bibitem{Bastianelli:2007pv}
  F.~Bastianelli, O.~Corradini and E.~Latini,
  ``Higher spin fields from a worldline perspective,''
  JHEP {\bf 0702} (2007) 072
  [hep-th/0701055].

\bibitem{Bastianelli:2008nm}
  F.~Bastianelli, O.~Corradini and E.~Latini,
  ``Spinning particles and higher spin fields on (A)dS backgrounds,''
  JHEP {\bf 0811} (2008) 054
  [arXiv:0810.0188 [hep-th]].

\bibitem{Bastianelli:2012bn}
  F.~Bastianelli, R.~Bonezzi, O.~Corradini and E.~Latini,
  ``Effective action for higher spin fields on (A)dS backgrounds,''
  JHEP {\bf 1212} (2012) 113
  [arXiv:1210.4649 [hep-th]].

\bibitem{Bastianelli:2014lia}
  F.~Bastianelli, R.~Bonezzi, O.~Corradini and E.~Latini,
  ``Massive and massless higher spinning particles in odd dimensions,''
  JHEP {\bf 1409} (2014) 158
  [arXiv:1407.4950 [hep-th]].

\bibitem{Bastianelli:2002fv}
  F.~Bastianelli and A.~Zirotti,
  ``Worldline formalism in a gravitational background,''
  Nucl.\ Phys.\ B {\bf 642} (2002) 372
  [hep-th/0205182].

\bibitem{Brink:1976sz}
  L.~Brink, S.~Deser, B.~Zumino, P.~Di Vecchia and P.~S.~Howe,
  ``Local supersymmetry for spinning particles,''
  Phys.\ Lett.\ B {\bf 64} (1976) 435.

\bibitem{Henneaux:1982ma}
  M.~Henneaux and C.~Teitelboim,
  ``Relativistic quantum mechanics of supersymmetric particles,''
  Annals Phys.\  {\bf 143} (1982) 127.


\bibitem{Bastianelli:2002qw}
F.~Bastianelli, O.~Corradini and A.~Zirotti,
``Dimensional regularization for SUSY sigma models and the worldline formalism,''
Phys.\ Rev.\ D {\bf 67} (2003) 104009
[arXiv:hep-th/0211134];

\bibitem{Bastianelli:2003bg}
F.~Bastianelli, O.~Corradini and A.~Zirotti,
``BRST treatment of zero modes for the worldline formalism in curved space,''
JHEP {\bf 0401} (2004) 023
[arXiv:hep-th/0312064].

\bibitem{Howe:1989vn}
  P.~S.~Howe, S.~Penati, M.~Pernici and P.~K.~Townsend,
  ``A particle mechanics description of antisymmetric tensor fields,''
  Class.\ Quant.\ Grav.\  {\bf 6} (1989) 1125.

\bibitem{Bastianelli:2005vk}
  F.~Bastianelli, P.~Benincasa and S.~Giombi,
  ``Worldline approach to vector and antisymmetric tensor fields,''
  JHEP {\bf 0504} (2005) 010
  [hep-th/0503155].

\bibitem{Bastianelli:2005uy}
  F.~Bastianelli, P.~Benincasa and S.~Giombi,
  ``Worldline approach to vector and antisymmetric tensor fields. II.,''
  JHEP {\bf 0510} (2005) 114
  [hep-th/0510010].

\bibitem{Bargmann:1948ck}
  V.~Bargmann and E.~P.~Wigner,
  ``Group theoretical discussion of relativistic wave equations,''
  Proc.\ Nat.\ Acad.\ Sci.\  {\bf 34} (1948) 211.

\bibitem{Bekaert:2002dt}
  X.~Bekaert and N.~Boulanger,
  ``Tensor gauge fields in arbitrary representations of GL(D,R): Duality and Poincare lemma,''
  Commun.\ Math.\ Phys.\  {\bf 245} (2004) 27
  [hep-th/0208058].

\bibitem{Bekaert:2003az}
  X.~Bekaert and N.~Boulanger,
  ``On geometric equations and duality for free higher spins,''
  Phys.\ Lett.\ B {\bf 561} (2003) 183
  [hep-th/0301243].

\bibitem{Bekaert:2003zq}
  X.~Bekaert and N.~Boulanger,
  ``Mixed symmetry gauge fields in a flat background,''
  hep-th/0310209.

\bibitem{Bekaert:2006ix}
  X.~Bekaert and N.~Boulanger,
  ``Tensor gauge fields in arbitrary representations of GL(D,R). II. Quadratic actions,''
  Commun.\ Math.\ Phys.\  {\bf 271} (2007) 723
  [hep-th/0606198].

\bibitem{Siegel:1988ru}
  W.~Siegel,
  ``Conformal invariance of extended spinning particle mechanics,''
  Int.\ J.\ Mod.\ Phys.\ A {\bf 3} (1988) 2713.

\bibitem{Siegel:1988gd}
  W.~Siegel,
  ``All free conformal representations in all dimensions,''
  Int.\ J.\ Mod.\ Phys.\ A {\bf 4} (1989) 2015.

\bibitem{Fronsdal:1978rb}
  C.~Fronsdal,
  ``Massless fields with integer spin,''
  Phys.\ Rev.\ D {\bf 18} (1978) 3624.

\bibitem{Labastida:1987kw}
  J.~M.~F.~Labastida,
  ``Massless particles in arbitrary representations of the Lorentz group,''
  Nucl.\ Phys.\ B {\bf 322} (1989) 185.

\bibitem{Francia:2002pt}
  D.~Francia and A.~Sagnotti,
  ``On the geometry of higher spin gauge fields,''
  Class.\ Quant.\ Grav.\  {\bf 20} (2003) S473
  [hep-th/0212185].

\bibitem{Corradini:2010ia}
  O.~Corradini,
  ``Half-integer higher spin fields in (A)dS from spinning particle models,''
  JHEP {\bf 1009} (2010) 113
  [arXiv:1006.4452 [hep-th]].

\bibitem{Marnelius:1988ab}
  R.~Marnelius and U.~Martensson,
  ``BRST quantization of free massless relativistic particles of arbitrary spin,"
   Nucl.\ Phys.\  B {\bf 321} (1989) 185.

\bibitem{Siegel:1999ew}
  W.~Siegel,
  ``Fields,'' chapter XII,
  arXiv:hep-th/9912205.

\bibitem{Henneaux:1987cp}
  M.~Henneaux and C.~Teitelboim,
  ``First and second quantized point particles of any spin,''
  In *Santiago 1987, Proceedings, Quantum mechanics of fundamental systems 2* 113-152.

\bibitem{Pashnev:1990cf}
  A.~I.~Pashnev and D.~P.~Sorokin,
  ``On n=4 super field description of relativistic spinning particle mechanics,''
  Phys.\ Lett.\ B {\bf 253} (1991) 301.

\bibitem{Bastianelli:2013pta}
  F.~Bastianelli, R.~Bonezzi, O.~Corradini and E.~Latini,
  ``Particles with non abelian charges,''
  JHEP {\bf 1310} (2013) 098
  [arXiv:1309.1608 [hep-th]].

\bibitem{Bastianelli:2013tsa}
  F.~Bastianelli and R.~Bonezzi,
  ``One-loop quantum gravity from a worldline viewpoint,''
  JHEP {\bf 1307} (2013) 016
  [arXiv:1304.7135 [hep-th]].

\bibitem{Marcus:1994mm}
  N.~Marcus,
 ``Kahler spinning particles,''
  Nucl.\ Phys.\ B {\bf 439} (1995) 583
  [hep-th/9409175].

\bibitem{Bastianelli:2009vj}
  F.~Bastianelli and R.~Bonezzi,
  ``U(N) spinning particles and higher spin equations on complex manifolds,''
  JHEP {\bf 0903} (2009) 063
  [arXiv:0901.2311 [hep-th]].

\bibitem{Bastianelli:2011pe}
  F.~Bastianelli and R.~Bonezzi,
  ``Quantum theory of massless (p,0)-forms,''
  JHEP {\bf 1109} (2011) 018
  [arXiv:1107.3661 [hep-th]].

\bibitem{Bastianelli:2012nh}
  F.~Bastianelli, R.~Bonezzi and C.~Iazeolla,
  ``Quantum theories of (p,q)-forms,''
  JHEP {\bf 1208} (2012) 045
  [arXiv:1204.5954 [hep-th]].

\bibitem{Campoleoni:2012th}
  A.~Campoleoni and D.~Francia,
  ``Maxwell-like Lagrangians for higher spins,''
  JHEP {\bf 1303} (2013) 168
  [arXiv:1206.5877 [hep-th]].

\bibitem{Francia:2010qp}
  D.~Francia,
  ``String theory triplets and higher-spin curvatures,''
  Phys.\ Lett.\ B {\bf 690} (2010) 90
  [arXiv:1001.5003 [hep-th]].

\bibitem{Francia:2012rg}
  D.~Francia,
  ``Generalised connections and higher-spin equations,''
  Class.\ Quant.\ Grav.\  {\bf 29} (2012) 245003
  [arXiv:1209.4885 [hep-th]].

\bibitem{Bekaert:2015fwa}
  X.~Bekaert, N.~Boulanger and D.~Francia,
  ``Mixed-symmetry multiplets and higher-spin curvatures,''
  arXiv:1501.02462 [hep-th].

\bibitem{Scherk:1979zr}
  J.~Scherk and J.~H.~Schwarz,
  ``How to get masses from extra dimensions,''
  Nucl.\ Phys.\ B {\bf 153} (1979) 61.

\bibitem{Hallowell:2007zb}
  K.~Hallowell and A.~Waldron,
  ``The symmetric tensor Lichnerowicz algebra and a novel associative Fourier-\
-Jacobi algebra,''
  SIGMA {\bf 3} (2007) 089
  [arXiv:0707.3164 [math.DG]].

\bibitem{Bastianelli:2009eh}
  F.~Bastianelli, O.~Corradini and A.~Waldron,
  ``Detours and paths: BRST complexes and worldline formalism,''
  JHEP {\bf 0905} (2009) 017
  [arXiv:0902.0530 [hep-th]].

\bibitem{Kuzenko:1995mg}
  S.~M.~Kuzenko and Z.~V.~Yarevskaya,
  ``Conformal invariance, N extended supersymmetry and massless spinning particles in anti-de Sitter space,''
  Mod.\ Phys.\ Lett.\ A {\bf 11} (1996) 1653
  [hep-th/9512115].

\bibitem{Bastianelli:2011cc}
  F.~Bastianelli, R.~Bonezzi, O.~Corradini and E.~Latini,
  ``Extended SUSY quantum mechanics: transition amplitudes and path integrals,''
  JHEP {\bf 1106} (2011) 023
  [arXiv:1103.3993 [hep-th]].

\bibitem{Bastianelli:2000dw}
  F.~Bastianelli and O.~Corradini,
  ``6-D trace anomalies from quantum mechanical path integrals,''
  Phys.\ Rev.\ D {\bf 63} (2001) 065005
  [hep-th/0010118].

\bibitem{Bastianelli:2001tb}
  F.~Bastianelli and N.~D.~Hari Dass,
  ``Simplified method for trace anomaly calculations in $d \leq6$,''
  Phys.\ Rev.\ D {\bf 64} (2001) 047701
  [hep-th/0104234].

\end{thebibliography}


\end{document}